\documentstyle[prl,aps,epsf]{revtex} \def\narrowtext{} \tighten \twocolumn
\input epsf.sty 
\begin{document}
\draft
 
\title{Electron Self-Energy of High Temperature Superconductors as Revealed
by Angle Resolved Photoemission}
\author{
        M. R. Norman,$^1$
        H. Ding,$^{1,2}$
        M. Randeria,$^3$
        and J. C. Campuzano,$^{1,2}$
       }
\address{
         (1) Materials Sciences Division, Argonne National Laboratory,
             Argonne, IL 60439 \\
         (2) Department of Physics, University of Illinois at Chicago,
             Chicago, IL 60607\\
         (3) Tata Institute of Fundamental Research, Bombay 400005, India\\
         }

\address{%
\begin{minipage}[t]{6.0in}
\begin{abstract}
In this paper, we review some of the work our group has done in the past few
years to obtain the electron self-energy of high temperature superconductors
by analysis of angle-resolved photoemission data.  We focus on three examples
which have revealed:  (1) a d-wave superconducting gap, (2) a collective
mode in the superconducting state, and (3) pairing correlations in the
pseudogap phase.  In each case, although a novel result is obtained which
captures the essense of the data, the conventional physics used leads to an
incomplete picture.  This indicates that new physics needs to be developed to
obtain a proper understanding of these materials.
\typeout{polish abstract}
\end{abstract}
\end{minipage}}

\maketitle
\narrowtext

Eleven years after their discovery, the physics of high temperature
superconductors is still not well understood because of their complex nature.
One of the key tools used to obtain information on these materials has been
angle resolved photoemission spectroscopy (ARPES).  Although a
surface sensitive probe, ARPES has the advantage of being resolved both in 
energy and momentum space, thus providing information difficult to
obtain from other methods.  Given the fact that ARPES measures
the single particle spectral function \cite{MOHIT}, then, in principle, one 
should be able to obtain the electron self-energy from the data.  
In some sense, this would ``solve" the high temperature superconductor problem,
assuming one had some microscopic theory which produced the same self-energy.

In this spirit, our group has worked several years now analyzing ARPES data
in an attempt to extract useful information about the electron self-energy for
high temperature cuprate superconductors.  The amount of work done is too 
extensive to review in this short paper, so we will confine ourselves to three
examples.  In each case we find a non-trivial result
which captures the essense of the data.  But in each case, we find that our
``conventional" explanation is in some sense incomplete.  We will use this
to show that any ``mean field" explanation of the data will always lead to
inconsistencies and relate this to the long standing ``x" versus ``1+x" 
debate on the doping dependence of physical quantities.
The conclusion is that new physics will need to be developed to obtain a
complete picture of the data.

Our first example concerns the determination of the low temperature
superconducting gap. Traditionally, workers in ARPES\cite{SHEN93} have
defined the gap
by the midpoint of the leading edge of the spectrum. Although this midpoint
is related to the superconducting gap, it is not the same\cite{ROLAND}.
At low temperatures, and ignoring linewidth broadening and momentum resolution,
the midpoint of spectra at the Fermi momentum is the superconducting gap
minus the HWHM of the energy resolution, if the gap is large enough so that
the Fermi function plays no role.  Even with these restrictions, this
statement assumes one can equate the photocurrent to the spectral function,
and that one knows the Fermi momentum, each of which involves a number of
assumptions.  We have taken the first step beyond this midpoint 
criterium in an attempt to give a well-defined meaning to the 
measurement of the gap by ARPES \cite{PRL95}.

We first assert that the measured photocurrent is proportional to the spectral
function times the Fermi function, the proportionality constant being the
dipole matrix element connecting the initial and final states (the signal
above the Fermi energy, due to higher harmonics of the photon beam,
is obviously subtracted before making this identification).  This assumes
(1) the sudden approximation is valid, (2) contributions due to the gradient
of the photon vector potential can be ignored, and (3) ``secondaries" (due
to inelastic scattering of the photoelectron) are either small or have 
also been
subtracted.  Although this seems a lot to stomach at once, there are ways
to test this.  For instance, if valid, then a frequency integral of the ARPES
spectra should be proportional to the momentum distribution function,
$n_{\bf k}$.  Our studies\cite{MOHIT} indeed indicate that the
frequency integrated ARPES data are consistent with such an identification.
Exploiting this, a rigorous method can be suggested to determine the Fermi
momentum, that point where the gradient of the integrated data
(i.e., $|\nabla n_{\bf k}|$) has a maximum\cite{PHMIX}.  Doing this, we find a
large hole-like Fermi surface centered about the $(\pi,\pi)$ points of the
square lattice Brillouin zone with a volume consistent with 1+x\cite{PRL96}.

We next need some model spectral function by which to fit the data.  Since we
wish to determine the BCS gap, then it is natural to use a BCS spectral
function.  At low temperatures we are fortunate, in that the
leading edge of the spectrum is resolution limited.  This implies that the
imaginary part of the electron self-energy is small at frequencies of order
the superconducting gap, and thus there is some justification for using a
BCS-like ansatz (limitations of this picture will be discussed in our next
example).  For frequencies beyond the gap, though, self-energy effects cannot
be ignored.  One sees this in the data as a large non-quasiparticle peak
contribution to the spectrum.  Whether all of this weight is part of the true
spectral function or not (i.e., secondaries) is still a matter of debate.
In our case, we simply subtracted this
incoherent part off by assuming it could be treated as ``secondaries" (using
the standard Shirley procedure).  In
practice, for extracting the gap, this makes little difference since one is
fitting only the leading edge of the spectrum, which is not much affected by
this subtraction (though this subtraction 
becomes more of an issue as the temperature is
raised).  But this does illustrate the point made at the beginning.  Although by
doing this subtraction we are able to fit the data to a BCS spectral function,
and thus obtain a reliable estimate of the BCS gap, we have in essence only
obtained a single number.  Although it is a very useful number,
as we
will see, it encapsulates only one aspect of the very interesting ARPES spectra
in the superconducting state.

\begin{figure}
\vspace{0.5cm}
\epsfysize=3.0in
\epsfbox{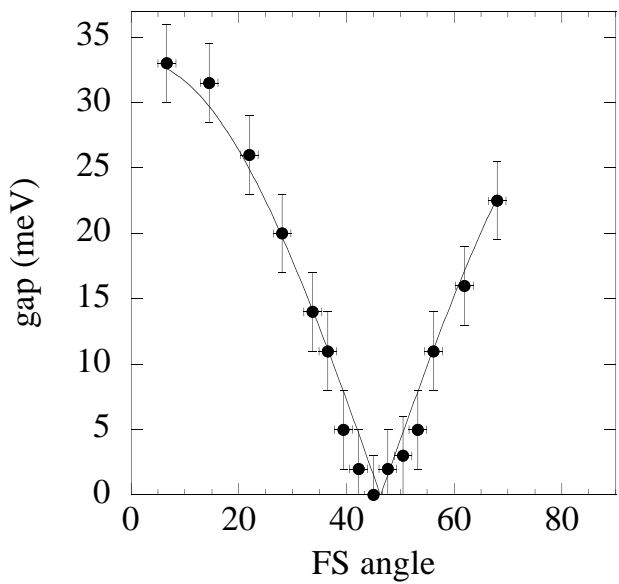}
\vspace{0.5cm}
\caption{$Y$ quadrant gap in meV versus angle on the Fermi surface
(filled circles) compared to a $d_{x^2-y^2}$ gap (solid curve).}
\label{fig1}
\end{figure}

Our first attempt at this procedure revealed a gap in Bi2212 which was a strong
function of
the Fermi momentum\cite{PRL95}, in support of earlier work by the group of
Z.-X. Shen\cite{SHEN93}.  Unlike this earlier work, evidence for two zeros
of the gap as a function of momentum (per zone quadrant), rather than the
single zero expected for a d-wave order parameter, was found.
At that time, we suggested two possibilities for interpreting this:  (1) an
anisotropic s-wave gap, or (2) a d-wave gap which was either being measured
on the true Fermi surface or one of the ghost images of the Fermi surface
(the ghosts due to diffraction of the outgoing photoelectrons by the
incommensurate BiO superlattice) depending on what particular value of momentum 
one was measuring.
Subsequently, by exploiting the photon polarization dependence of the dipole
matrix elements, we were able to show that explanation (2) was actually the
correct one\cite{NORM95,GAP96}. As extensively discussed in these papers, the
superlattice complications can be avoided by measuring the gap in the Y
quadrant of the Brillouin zone. Doing so reveals a gap which beautifully
follows the behavior predicted for an order parameter with $d_{x^2-y^2}$
symmetry\cite{GAP96} (see Fig.~1). In fact, one learns more than this.
Since the data follow the form $\cos(k_xa)-\cos(k_ya)$ quite closely, this
indicates that the pairing interaction is fairly local in real space.  Data
taken on another optimally doped sample where the large gap region was 
sampled more closely\cite{GAP96} actually indicate the presence of a weak
maximum in the gap at locations on the Fermi surface connected by $(\pi,\pi)$ 
wavevectors.  Similar effects have been seen in calculations where spin
fluctuations are considered as the pairing mechanism.

As said above, even though a lot of useful information is obtained from 
knowing the value of $\Delta_{\bf k}$, it is only a small part of the 
overall story.  This can be seen in Fig.~2, where the spectrum at the $(\pi,0)$ 
point of the zone for a slightly overdoped sample at low temperatures is shown.
Rather interestingly, this spectrum agrees with that of the normal state
for energies beyond about 90 meV, which is equivalent to stating that
the self-energies agree beyond this energy.  For lower energies, though, 
one sees a dramatic departure of the superconducting state spectra from 
the normal state one, as first noted by Dessau {\it et al.}\cite{DESSAU}.  The
superconducting (SC) 
state spectrum first drops (thus leading to a dip/hump structure) then 
rises to form a sharp, essentially resolution limited, quasiparticle 
peak.  Since this change in behavior is all occuring on the scale of the energy 
resolution, this indicates that the imaginary part of the 
self-energy ($Im\Sigma$) must drop from its large normal state value to a 
small value over a narrow energy range.  Fits we have done using 
model self-energies reveal that the drop in $Im\Sigma$ must be rather 
abrupt, essentially a step edge (the standard d-wave prediction of crossing
over from $\omega$ to $\omega^3$ is too weak to give a dip). In fact, 
the observed dip is so deep, it is best fit by a peak in $Im\Sigma$ 
followed by a rapid drop.

There are a number of consequences of such behavior.  By 
Kramers-Kronig transformation, a step in $Im\Sigma$ implies a peak in 
$Re\Sigma$.  Such a peak will lead to an additional mass renormalization 
relative to the normal state which acts to suppress the quasiparticle 
dispersion.  This explains the rather puzzling observation that the 
quasiparticle peak does not appear to disperse much when moving away from 
the $(\pi,0)$ point. In fact, data on a number of our samples indicate 
that a sharp low energy peak is still present when moving towards the $(0,0)$ 
point even when the higher binding energy feature (the hump) has begun to 
disperse (the hump dispersion rapidly approaches the dispersion of the 
single broad peak seen in the normal state).  This result was so puzzling
that the data were not published for several years. The peak in $Re\Sigma$ 
however naturally explains this, since it can lead to a low energy 
quasiparticle pole even when the normal state binding energy moves well away 
from the Fermi energy.  We 
note that a step edge in $Im\Sigma$ is equivalent to the problem of an 
electron interacting with a dispersionless mode, as previously treated 
by Engelsberg and Schrieffer \onlinecite{SCHR}. Such an 
interaction leads to the prediction of a spectral function composed of 
two features (peak and hump) whose dispersion is remarkably similar to 
that extracted from the data\cite{NORM97a}. The crucial difference here 
is that this behavior appears only in the superconducting state, and thus 
the mode is not a phonon. Rather, it must be of collective origin.
Detailed calculations we have done of a superconducting electron interacting
with a dispersionless mode give a good description of the 
data\cite{NORM97b}, with the peak in $Im\Sigma$ due 
to the peak in the SC density of states. They also give a good fit to 
the observed dispersion of the two features.

\begin{figure}
\vspace{0.5cm}
\epsfysize=3.0in
\epsfbox{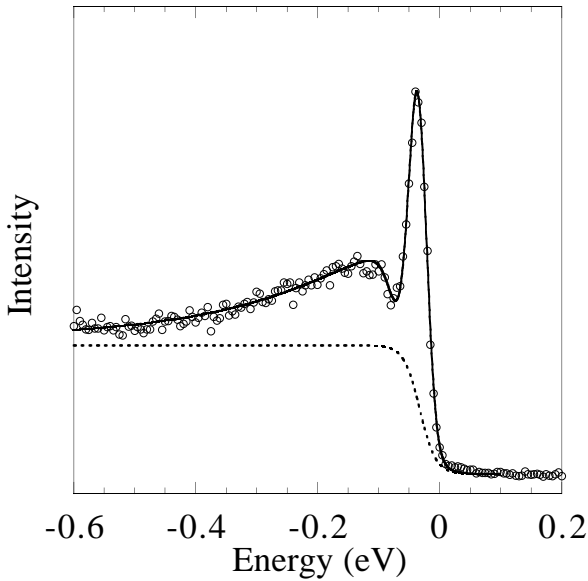}
\vspace{0.5cm}
\caption{
Comparison of SC data at $(\pi,0)$ to a model fit based on electrons
interacting with a collective mode.  The dashed line is an assumed
background contribution.}
\label{fig2}
\end{figure}

Microscopically, these findings imply that the dominant contribution 
to the electron self-energy is from electron-electron scattering 
processes. The low frequency reduction in $Im\Sigma$ is a consequence of 
the gapping of the spectrum causing the scattering (i.e., $\alpha^2F$) by the
superconducting gap (the 2$\Delta$ effect first discussed by Kuroda and
Varma\cite{VARMA}).  To get a step edge, though, one must 
assume that the gapped weight shows up as a sharp mode inside 
of this ``2$\Delta$" gap.  Recent spin fluctuation theories have indeed 
predicted such behavior\cite{BICKERS}.  Fits to the data 
indicate a mode with an energy of 41 meV, equivalent to that of the 
collective mode seen by neutron scattering in YBCO\cite{YBCO}.  Whether 
this is a conincidence or not remains to be seen. The YBCO neutron 
scattering data indicate 
that the mode is associated with $(\pi,\pi)$ scattering events, whereas 
in the ARPES data
the step edge in $Im\Sigma$ implies dipsersionless behavior.  On the 
other hand, the low energy ARPES peak exists over about the same momentum range 
along $(\pi,0)-(\pi,\pi)$ as it does along $(\pi,0)-(0,0)$.  As these 
directions are related by a $(\pi,\pi)$ translation, the ARPES data 
also indicate that 
$(\pi,\pi)$ scattering is indeed playing an important role. This is most 
obviously seen in the fact that the dip/hump structure is most pronounced 
in spectra at the $(\pi,0)$ points, as recently emphasized by Shen and 
Schrieffer\cite{SS}.

It is important to remark that the above description is incomplete. 
In Fig.~2, we show that 
our model gives a very good fit to the spectra, but at a price. The 
price is that a large ``background" contribution has to be subtracted 
off the data. This background is modeled by a step, and is essentially 
equivalent to the total ARPES spectra for unoccupied states (with the 
step edge at the Fermi energy in the normal state, but pushed back by 
$\Delta$ in the superconducting state). In reality, there are 
indications that most if not all of 
this background is part of the true spectral function.  This has led to a recent
speculation that this large background actually represents the 
instability of the photohole to decay into spinons and 
holons\cite{LAUGH}. In such a model, the quasiparticle peak represents a 
bound state split from this continuum.  To look into these matters in more
detail, we have recently attempted to extract the 
actual experimental self-energy by direct inversion of the ARPES 
data.  This inversion reveals the predicted peak and step edge 
in $Im\Sigma$ of our model, as well as the peak in 
$Re\Sigma$\cite{SELF}.

We now ask ourselves the question of how the above picture changes as 
the doping is reduced, moving towards the Mott insulating phase.  What is 
found is that the low temperature gap again has the expected d-wave
form\cite{NATURE}. On the other hand, the spectrum does change 
significantly as the doping is reduced. The quasiparticle peak becomes smaller 
and the hump becomes more pronounced, moving to higher binding energy.  
Perhaps the most signficant change is seen upon heating the sample. In 
overdoped Bi2212, the superconducting gap is observed to close at or near 
$T_c$.  Unpublished fits we have done similar to those of Fig.~1 (which are
obviously more suspect as the temperture is raised) reveal a gap whose T 
dependence is much flatter than the BCS prediction, which then closes 
rapidly near $T_c$.  This is accompanied by a strong increase in the low 
frequency broadening back to its large normal state value (the broadening
is found to drop approximately like $T^6$ below $T_c$, similar to what is 
seen in conductivity meaurements, and again a strong indication of the 
electron-electron scattering origin of $\Sigma$).  In the underdoped 
case, however, something quite different occurs.  The spectral gap is seen 
to exist well 
above $T_c$, only disappearing at a higher temperature (denoted $T^*$) 
\cite{MARSHALL,NATURE,LOESER} and has a similar anisotropy as that seen 
below $T_c$.  The gap as measured by the midpoint of the leading edge 
goes smoothly through $T_c$\cite{NATURE} indicating that the gap above 
$T_c$ has the same origin as the gap below $T_c$, as predicted by theories
with pairing correlations above $T_c$\cite{MOH1}.

This so-called pseudogap has been seen in a variety of other measurements, 
most of them predating the ARPES ones.  Its origin is a matter of intense 
debate and encapsulates one of the most fundamental issues of high 
temperature superconductivity:  how the unusual superconducting state 
seen in the cuprates evolves into the equally unusual Mott insulating 
state.  Again, the advantage of ARPES is that it provides both momentum and 
frequency resolved information.  
What does it find?  First, the quasiparticle peaks appear only 
below $T_c$ (i.e., not below $T^*$), and again one finds that the low 
frequency broadening drops off roughly as $T^6$ below $T_c$.  Above 
$T_c$, the spectra near $(\pi,0)$ are quite unusual, being rather flat, 
but with a sharp leading edge with a large gap
(note, this leading edge gap discussed in 
Refs.\onlinecite{NATURE,LOESER} is not the same as the hump position 
discussed in Ref.\onlinecite{MARSHALL}, a point of confusion in the 
literature).  Second, this leading edge gap, as characterized by its midpoint, 
smoothly evolves through $T_c$\cite{NATURE,NEW} implying that it is of 
the same origin as the superconducting gap.  Third, the low temperature
gap actually increases as the doping decreases, reflecting the increase 
of $T^*$ with underdoping\cite{HARRIS,MIYA,NEW}, again showing the 
strong connection between the superconducting gap and pseudogap.  
Fourth, the pseudogap has a similar anisotropy above $T_c$ to that
below\cite{NATURE,LOESER}.  Fifth, the pseudogap is tied to the normal 
state Fermi surface\cite{PRL97}, as expected if the gap were of pairing 
origin, rather than of CDW or SDW origin.

Recently, we have looked again into the anisotropy issue by taking 
temperature sweeps at different ${\bf k}$ points on the Fermi surface.  
We found somthing quite unusual, in that the pseudogap disappears at 
different temperatures for different ${\bf k}$\cite{NEW}.  This means 
that the d-wave node below $T_c$ becomes a gapless arc above $T_c$  
which expands in 
length with temperature, until the entire Fermi surface is recovered at $T^*$.  
The evolution is smooth, as predicted by theories with d-wave pairing
correlations\cite{MOH2}, rather than the abrupt change one would expect if 
the gap above $T_c$ were of different origin than the one below $T_c$.

Closer inspection, though, reveals that more is going on.  This is most 
clearly seen by employing a recent method we have suggested for removing 
the effects of the Fermi function from ARPES data\cite{NEW}.  If one 
assumes the spectral function is particle-hole symmetric, a mild 
assumption for spectra at $k_F$ over a sufficiently narrow energy range 
about the Fermi energy, then one can formally eliminate the Fermi 
function from the data by summing the ARPES intensity at positive and 
negative energies (with respect to the chemical potential).  By doing so, 
one acquires a dramatic visual picture of what is going on.  Such 
symmetrized data are shown in Fig.~3 at the $(\pi,0)-(\pi,\pi)$ Fermi crossing
for a moderately underdoped ($T_c$=83K) sample.  One clearly sees from 
this that the spectral gap fills in rather than closes.  In contrast, 
halfway along the Fermi surface between $(\pi,0)$ to $(\pi,\pi)$ the 
behavior is quite different, as the gap is seen to actually close (at a 
temperature between $T_c$ and $T^*$) rather than fill in.  The momentum 
dependence of the electron self-energy is highly non-trivial.

\begin{figure}
\vspace{0.5cm}
\epsfysize=3.0in
\epsfbox{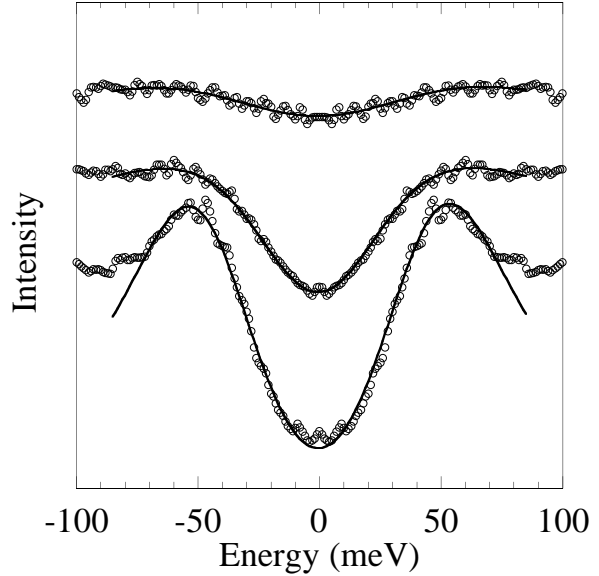}
\vspace{0.5cm}
\caption{
Symmetrized data on a $T_c$=83K sample at the $(\pi,0)-(\pi,\pi)$ Fermi
crossing at three temperatures (14K, 90K, and 170K) compared to
model fits based on pairing fluctuations.}
\label{fig3}
\end{figure}

These results have motivated us to find a model self-energy which captures 
the unusual behavior seen near the $(\pi,0)$ point in the 
pseudogap phase.  One which reproduces the low energy
data quite well (see Fig.~3) is of the form
$-i\Gamma_1+\Delta^2/(\omega+i\Gamma_0)$\cite{NEWb}.
BCS theory is recovered by setting 
$\Gamma_0$ to zero.  The surprising finding is that $\Gamma_0$ is 
proportional to $T-T_c$ (and thus zero below $T_c$), with $\Delta$
essentially T independent (the 
latter having been infered earlier from specific heat data in 
YBCO\cite{LORAM}).  $T^*$ then corresponds to where $\Delta=\Gamma_0(T)$, 
that is, although the spectral gap closes, $\Delta$ is still non-zero.  
The interesting point is that there is only one known quantity which is 
proportional to $T-T_c$, the inverse Cooper pair lifetime\cite{ELIHU}.  
This almost certainly means that the pseudogap is due to pairing 
correlations\cite{NEWb}.  In CDW or SDW type theories, the quantity 
$T-T_c$ would not naturally arise.  In fact,
the above form for the self-energy can be motivated by a t-matrix 
calculation of the self-energy due to pairing fluctuations\cite{NEWb}.  
The derivation is especially transparent in the limit where the bands are 
dispersionless, which thus motivates why the behavior is seen only near 
the $(\pi,0)$ points of the zone.

This ``zero dimensional" behavior again emphasizes the unconventional 
nature of the cuprates. A model having these characteristics has been 
recently proposed by Geshkenbein, Ioffe, and Larkin\cite{GESH}, who 
emphasized that such behavior can explain the existence of large pairing 
correlations without the large expected signatures of fluctuational conductivity
or diamagnetism. In essense, the states near $(\pi,0)$ have no measurable Fermi 
velocity and not even a remnant of a quasiparticle peak in the pseudogap
phase, and consequently they do not contribute to the supercurrent response. 
Any type of mean field treatment would therefore be totally inadequate.    
Although these electrons near $(\pi,0)$  have a large $\Delta$, they have 
no $\Psi$ in the Ginzburg-Landau sense.  
As the doping decreases, the anomalous 
region expands, eventually taking over the entire zone, giving rise to 
the non-superconducting, insulating state (data taken on low $T_c$ Bi2212 
samples show this type of behavior\cite{NATURE,PRL97}).

Any mean field description of the above would force 
one again into the two gap picture, which is inconsistent with the data, 
in that the gap for each {\bf k} smoothly evolves through $T_c$, and at low
temperatures has an anisotropy completely consistent with a simple d-wave 
order parameter.  This same kind of inconsistency is also seen if one 
attempts to force a Fermi liquid picture in the underdoped 
superconducting state\cite{LEE}, and is related to the ``x" versus 
``1+x" debate which has been prevalent in the cuprate literature over 
the past decade, a debate which encapsulates the 
unconventional nature of the cuprates.  Is it x, or 1+x?  In some sense, 
the ARPES data say it is both, analogous to quantum mechanics whose objects
behave like waves or particles depending on what question one is asking.
Is there a large Fermi surface enclosing 
a volume 1+x, even for reduced doping?  The answer is yes, 
but\cite{PRL97}.  The ``but" is due to the fact that one can quite 
reasonably define a Fermi crossing along $(\pi,0)-(\pi,\pi)$ due to the 
intensity drop in the spectra (the $|\nabla n_{\bf k}|$ maximal argument 
discussed earlier), but this hides the fact that there is no true 
dispersion in the conventional band theory sense.  In essence, states 
near the $(\pi,0)$ points do not behave like a normal Fermi liquid, and 
the rest form a liquid whose effective volume is incresingly reduced 
as the doping is reduced, leading to x-like scaling (note, this is not 
the same as the ``mean field" x picture with small hole pockets, something
we find no evidence for).  This behavior is 
perhaps most exhaustively discussed in the recent book of 
Anderson\cite{AND}.

In conclusion, much useful information concerning the electron 
self-energy can be extracted from an analysis of the ARPES data.  In each 
example studied above, a conventional, but exotic, explanation is found 
(d-wave superconducting gap, collective mode in the superconducting 
state, pairing correlations above $T_c$).  But, in each case, the 
conventional explanation, although capturing the essence of the data, is 
incomplete.  This indicates that new physics
needs to be developed before a true understanding of the cuprates is achieved. 

This work was supported by the U. S. Dept. of Energy,
Basic Energy Sciences, under contract W-31-109-ENG-38, the National 
Science Foundation DMR 9624048, and
DMR 91-20000 through the Science and Technology Center for Superconductivity.

\end{document}